\documentclass[AMA,STIX1COL]{WileyNJD-v2}
\usepackage{draftwatermark}

\SetWatermarkText{ACCEPTED VERSION}
\SetWatermarkScale{.5}
\SetWatermarkAngle{45}

\articletype{Tutorial in Biostatistics}%

\received{26 April 2016}
\revised{6 June 2016}
\accepted{6 June 2016}

\begin{document}

\title{Correcting for partial verification bias in diagnostic accuracy studies: A tutorial using R\thanks{This is the accepted version of the following article: Arifin WN and Yusof UK. Correcting for partial verification bias in diagnostic accuracy studies: A tutorial using R. Statistics in Medicine. 2022;41(9):1709-1727, which has been published in final form at https://onlinelibrary.wiley.com/doi/10.1002/sim.9311. This article may be used for non-commercial purposes in accordance with the Wiley Self-Archiving Policy [http://www.wileyauthors.com/self-archiving].}}

\author[1,2]{Wan Nor Arifin}

\author[1]{Umi Kalsom Yusof*}

\authormark{ARIFIN \textsc{et al}}

\address[1]{\orgdiv{School of Computer Sciences}, \orgname{Universiti Sains Malaysia}, \orgaddress{\state{Pulau Pinang}, \country{Malaysia}}}

\address[2]{\orgdiv{Biostatistics and Research Methodology Unit, School of Medical Sciences}, \orgname{Universiti Sains Malaysia}, \orgaddress{\state{Kelantan}, \country{Malaysia}}}

\corres{*Associate Professor Dr. Umi Kalsom Yusof, School of Computer Sciences, Universiti Sains Malaysia, 11800 Pulau Pinang, Malaysia. \email{umiyusof@usm.my}}

\abstract[Summary]{Diagnostic tests play a crucial role in medical care. Thus any new diagnostic tests must undergo a thorough evaluation. New diagnostic tests are evaluated in comparison with the respective gold standard tests. The performance of binary diagnostic tests is quantified by accuracy measures, with sensitivity and specificity being the most important measures. In any diagnostic accuracy study, the estimates of these measures are often biased owing to selective verification of the patients, which is referred to as partial verification bias. Several methods for correcting partial verification bias are available depending on the scale of the index test, target outcome and missing data mechanism. However, these are not easily accessible to the researchers due to the complexity of the methods. This article aims to provide a brief overview of the methods available to correct for partial verification bias involving a binary diagnostic test and provide a practical tutorial on how to implement the methods using the statistical programming language R.}


\jnlcitation{\cname{%
\author{Arifin WN} and 
\author{Yusof UK}} (\cyear{20XX}). 
\ctitle{Correcting for partial verification bias in diagnostic accuracy studies: A tutorial using R}. \cjournal{Statistics in Medicine}, \cvol{20XX;00:1--N}.\\ \\
\thanks{This is the accepted version of the following article: Arifin WN and Yusof UK. Correcting for partial verification bias in diagnostic accuracy studies: A tutorial using R. Statistics in Medicine. 2022;41(9):1709-1727, which has been published in final form at https://onlinelibrary.wiley.com/doi/10.1002/sim.9311. This article may be used for non-commercial purposes in accordance with the Wiley Self-Archiving Policy [http://www.wileyauthors.com/self-archiving].}}

\maketitle


\section{Introduction}\label{sec1}
Evaluating a new diagnostic test (index test) involves comparing the index test results with the results of a definitive gold standard.\cite{zhou2011, pepe2011, zou2012, degroot2011a, osullivan2018} Diagnostic tests play a critical role in medical care in discriminating between patients with or without the disease. Therefore, any new diagnostic tests must undergo a thorough evaluation before being applied in clinical settings.\cite{kosinski2003a, osullivan2018, chikere2019}

The performance of a new diagnostic test is quantified by a number of accuracy measures, depending on the scale of the index test results.\cite{zhou2011, pepe2011} For binary index test results, sensitivity (true positive, i.e. the probability that the index test result is positive when the disease is present) and specificity (true negative, i.e. the probability that the index test result is negative when the disease is absent) are commonly used to indicate the accuracy.\cite{zhou2011, pepe2011, he2012, alonzo2014} Other commonly used accuracy measures are positive predictive value (PPV, the probability that the disease is present when the index test is positive) and negative predictive value (NPV, the probability that the disease is absent when the index test result is negative).\cite{zhou2011, pepe2011, linnet2012} For index test results in continuous scale, receiver operating characteristic (ROC) curve and area under the ROC curve are used to indicate the accuracy.\cite{zhou2011, pepe2011, zou2012, linnet2012, alonzo2014, chikere2019} In addition, the use of accuracy measures is not limited to the evaluation of diagnostic tests in medical research, but is also relevant to evaluate the predictive performance of machine (or statistical) learning methods.\cite{dhahri2019, li2019, sande2020, fleuren2020}

It is essential to obtain valid estimates of accuracy measures to ensure the clinical validity of the tests.\cite{chikere2019} Biased estimates are misleading, which may result in the premature implementation of the new tests and lead to wrong decision making by the clinicians.\cite{rutjes2007, degroot2011a} However, the assessment of index tests in diagnostic accuracy studies often suffer from bias because the verification of the target outcomes by gold standard tests can be clinically infeasible due to cost, ethical and clinical considerations, as well as time-consuming and invasive procedures.\cite{degroot2011a, alonzo2014, schmidt2015, osullivan2018}

In diagnostic accuracy studies, verification bias occurs when only a sub-sample of patients undergoing an index test is verified by the gold standard test based on the result of the index test.\cite{osullivan2018} Those with a positive index test result are more likely to be verified by the gold standard test because they are presumably more likely to be positive for the target outcome. In contrast, those with a negative index test result are less likely to be selected in the verification sample because they are presumably less likely to be positive for the target outcome. Selective sampling may also depend on other factors (covariates), such as gender, age and the presence of clinical symptoms.\cite{kosinski2003a, degroot2011b} Specifically, this selective sampling creates a partial verification bias, resulting in biased estimates of accuracy measures.\cite{rutjes2007, degroot2011a, osullivan2018}

Partial verification bias can be viewed as a missing data problem, where the target outcome is missing for a subset of the sample (unverified patients).\cite{degroot2011a, alonzo2014} In order to correct the bias, it is vital to know the missing data mechanism.\cite{alonzo2014} Missing completely at random (MCAR) occurs if the target outcome is missing regardless of the index test result and status of covariates. In this case there is no partial verification bias involved. In partial verification, there are two possible missing data mechanisms; missing at random (MAR) and missing not at random (MNAR). MAR happens if the target outcome is missing, depending only on the index test result and status of observed covariates. MNAR, also known as non-ignorable (NI), occurs when the target outcome is missing due to the target outcome itself (disease status) and the status of unobserved covariates, for example when the verification is withheld because of the disease status as determined by clinicians.\cite{kosinski2003b}

Several methods are available to correct for partial verification bias, depending on the scale of the index test, target outcome and missing data mechanism.\cite{alonzo2014, chikere2019} For interested readers, a list of the available methods is provided by a recent review by Chikere et al.\cite{chikere2019} This tutorial focuses on the binary index test (positive or negative) with a binary target outcome (disease or non-diseased). For diagnostic accuracy studies that suffer from partial verification bias, correcting the bias by utilising the available methods will improve the validity of the accuracy estimates. However, the methods are not easily accessible to the researchers due to the complexity of the methods.\cite{chikere2019} This article aims to provide a brief overview of the methods available to correct for partial verification bias involving a binary diagnostic test and provide a practical tutorial on how to implement the methods using the statistical programming language R.

\section{Correction methods for partial verification bias}\label{sec2}
This section gives a brief overview of the correction methods for partial verification bias involving binary diagnostic tests with binary outcome. The methods were categorised by the missingness mechanism that the methods can handle; MAR or MNAR.

\subsection{Data table}
A typical observed data table and notations are presented to describe the correction methods. Let's say all patients undergo a binary diagnostic test, and only a sub-sample of them undergo the gold standard test, the variables for an individual patient are\cite{begg1983}:

\begin{description}
\item[$T$,] the value of diagnostic test result: Binary = Positive (1), Negative (0)
\item[$D$,] the disease status as verified by gold standard test: Binary = Yes (1), No (0)
\item[$V$,] the verification status by gold standard test: Binary = Yes (1), No (0)
\item[$\mathbf{X}$,] the covariate(s): Discrete, continuous variable(s)
\end{description}

\noindent A typical table for this situation (without covariates)\cite{begg1983, kosinski2003a, harel2006} is given in Table \ref{tab:pvb}.

\begin{table*}[t]%
\caption{A typical data table when partial verification bias is present.\label{tab:pvb}}
\centering
\begin{tabular*}{500pt}{@{\extracolsep\fill}lcccc@{\extracolsep\fill}}
\toprule
&\multicolumn{2}{@{}c@{}}{\textbf{Disease $D = j$ verified, $V = 1$}} \\\cmidrule{2-3}
\textbf{Diagnostic test, $T$} & \textbf{$D = 1$}  & \textbf{$D = 0$}  & \textbf{Disease not verified, $V = 0$}  & \textbf{Total} \\
\midrule
$T = 1$ & $s_1$ & $r_1$ & $u_1$ & $n_1$\\
$T = 0$ & $s_0$ & $r_0$ & $u_0$ & $n_0$\\
\hline 
Total &  &  & $u$ & $n$\\
\bottomrule
\end{tabular*}
\begin{tablenotes}
\item Notations: $n$, the number of patients on which the new diagnostic test is studied; $n_1$, the number of patients who have a positive diagnostic test result; $n_0$, the number of patients who have a negative diagnostic test result; $u$, the number of patients who are not verified by the gold standard test; $u_1$, the number of patients who have a positive diagnostic test result, but are not verified by the gold standard test; $u_0$, the number of patients who have a negative diagnostic test result, but are not verified by the gold standard test; $s_1$, the number of patients with disease who have a positive diagnostic test result; $s_0$, the number of patients with disease who have a negative diagnostic test result; $r_1$, the number of patients without disease who have a positive diagnostic test result; $r_0$, the number of patients without disease who have a negative diagnostic test result.
\end{tablenotes}
\end{table*}

\subsection{MAR}
Many methods have been developed over the years to correct for partial verification bias. The first proposed method in this area of research is Begg and Greenes' method\cite{begg1983} (BG method), where they proposed a correction method based on Bayes' theorem. The method allows correction for bias with and without categorical covariates. In the presence of many categorical covariates, the method's correction is more problematic as it suffers from many combinations of $T$ and $\mathbf{X}$. Alonzo and Pepe\cite{alonzo2005} proposed that an extension to the BG method (EBG method) would allow the inclusion of continuous covariates. For correction involving only categorical covariates, the EBG method implemented by fitting a saturated logistic regression model is equivalent to the BG method. These methods are discussed further in the next section.

Alonzo and Pepe\cite{alonzo2005} also proposed three other methods to correct for partial verification bias, which are: mean score imputation (MSI), inverse probability weighting (IPW), and semi-parametric efficient (SPE) methods. The formula involved can be referred to in other sources.\cite{alonzo2005, he2012} The MSI method, proposed initially by Clayton et al.,\cite{clayton1998} was extended in the context of partial verification bias by Alonzo and Pepe.\cite{alonzo2005} The method combines the observed probability of disease given test result $P(D=1|T)$ with the predicted probability of disease given test result and covariates $\hat{P}(D=1|T,\mathbf{X})$ in calculating accuracy estimates, where the predicted probability is obtained from a logistic regression model. Alonzo and Pepe\cite{alonzo2005} derived the IPW correction, which was based on the work by Horvitz and Thompson\cite{horvitz1952}. The IPW method weighs each observation in the verified sample by the inverse of the probability of verification to provide the corrected sensitivity and specificity estimates.\cite{alonzo2005} Further, the estimated probability of verification $\hat{P}(V=1|T,\mathbf{X})$ is obtained from logistic regression.\cite{alonzo2005} Lastly, Gao et al.\cite{gao2000} and Alonzo et al.\cite{alonzo2003} independently derived the SPE estimator of disease prevalence for two-phase studies. Based on these studies,\cite{gao2000,alonzo2003} Alonzo and Pepe\cite{alonzo2005} derived bias-corrected estimators of sensitivity and specificity, where $\hat{P}(D=1|T,\mathbf{X})$ and $\hat{P}(V=1|T,\mathbf{X})$ required by the SPE method are obtained by logistic regression models. The SPE method is considered a semi-parametric method that is doubly robust because it is consistent when either the disease or verification model is incorrectly specified.\cite{alonzo2005, he2012}

Harel and Zhou\cite{harel2006} proposed using the multiple imputation (MI) method as a flexible and simulation-based alternative\cite{degroot2011a} to the BG method. It is flexible in such a way that it allows the use of imputation methods of choice. Although in its original form, it was proposed without covariates, de Groot et al.\cite{degroot2008, degroot2011a} and {\"U}nal and Burgut\cite{unal2014} showed that the MI method could be easily extended to include covariates. Another advantage of the MI method is that the MI procedure is readily available in most statistical software.\cite{degroot2011b} The method is discussed further in the next section.

A correction method based on the propensity score was proposed by He and McDermott.\cite{he2012} In the presence of many covariates, the dimension of covariates can be reduced by replacing it with propensity score, where the score can be estimated by using logistic regression or discriminant analysis.\cite{he2012} The propensity score is then used to stratify the subjects into subgroups, hence categorizing the covariate information into a single categorical variable.\cite{he2012}

\subsection{MNAR}
Following the landmark paper by Begg and Greenes\citep{begg1983} on the correction method for partial verification bias, Zhou\cite{zhou1993} extended the BG method to account for MNAR assumption. This method requires that the ratios of the conditional probability of verification for diseased to that of non-diseased when tested positive ($k_1$) and negative ($k_0$) for the index test are known. Because these ratios are usually unknown and difficult to estimate,\cite{kosinski2003a, alonzo2014} $k_1$ and $k_0$ are given as ranges of possible values, giving the lower and upper limits of sensitivity and specificity estimates.\cite{zhou1993}

Next, Kosinski and Barnhart\cite{kosinski2003a} proposed a likelihood-based logistic regression method to account for MNAR assumption. In this method, the maximum-likelihood estimates are obtained by the expectation-maximisation (EM) algorithm.\cite{kosinski2003a} The process of maximising the likelihood is simplified using the EM algorithm because the process is reduced to repeatedly fitting three logistic regression models in the maximisation step (M-step) and updating weights in the expectation step (E-step).\cite{kosinski2003a} The method is discussed further in the next section. In the same year, Kosinski and Barnhart\cite{kosinski2003b} also proposed a form of sensitivity analysis called global sensitivity analysis. This method provides a region of sensitivity and specificity values combinations consistent with the observed data, known as the test ignorance region. This region is then plotted to allow easy visualisation of the test ignorance region. Later, inspired by the EM-based method,\cite{kosinski2003a} {\"U}nal and Burgut\cite{unal2014} proposed using neural networks to replace logistic regression. They constructed the disease, diagnostic test and missing data mechanism components in three separate networks using the sigmoid function as the activation function instead of logistic regression models and using network weights rather than regression coefficients.\cite{unal2014} Rochani et al.\cite{rochani2015} proposed three log-linear models to handle MCAR, MAR and MNAR assumptions. The models are specified for a two-way table similar to Table \ref{tab:pvb}. Notably, the proposed models do not handle data with covariates.

Based on the Bayesian approach, four correction methods that handle both MAR and MNAR missing data mechanisms were proposed.\cite{martinez2006, buzoianu2008, pennello2011, hajivandi2018} The Bayesian approach differs from the other methods discussed so far as it includes prior information about the parameters of interest in inference. This information is elicited from expert opinion, previous research or other methods (for example by BG method).\cite{martinez2006, chikere2019} Bayesian-based methods overcome the model-identification problem and can handle significant number of parameters that the maximum-likelihood approach cannot handle.\cite{chikere2019, gelman2014} Martinez et al.\cite{martinez2006} extended Zhou's method for MNAR assumption\cite{zhou1993} to estimate the unknown $k_1$ and $k_0$ ratios, where these ratios are estimated by specifying Beta distribution priors for these quantities. The Markov Chain Monte Carlo (MCMC) procedure by Gibbs sampling algorithm was implemented, and the priors were obtained from the BG method.\citep{martinez2006} Buzoianu and Kadane\cite{buzoianu2008} provided the Bayesian version of the EM-based logistic regression method,\cite{kosinski2003a} where the data augmentation algorithm\cite{tanner1987, chikere2019} replaces the EM algorithm, and the method was implemented by Gibbs sampling. For models without covariates, Pennello\citep{pennello2011} specified three selection models: fully general NI (or MNAR), reduced NI and MAR models, and provided different prior specifications and Gibbs sampler designs for each model. The fully general NI model was specified with Beta distribution priors. In contrast, the reduced NI and MAR models were specified with non-informative priors for the relevant parameters in the models.\cite{pennello2011} Hajivandi et al.\cite{hajivandi2018} proposed a method for a specific situation where all those who tested negative on the index test are excluded from verification with the gold standard test. In this case, the information on index test result $T$ already gives complete verification information, thus the verification variable $V$ is redundant and eliminated from the conditional probability formulas to calculate sensitivity and specificity. The method \cite{hajivandi2018} only needs to obtain the posterior probability distribution of disease prevalence $P(D=1)$, and this parameter was estimated using the MCMC procedure using Adaptive Metropolis Block.

\section{Application in R}\label{sec3}

\subsection{Data set}\label{data_desc}
The single-photon-emission computed-tomography (SPECT) thallium test data set \citep{cecil1996, kosinski2003a} was used for the analysis in this tutorial. The SPECT thallium test is a non-invasive test utilised to diagnose coronary artery disease (CAD), which is the index test in this case. CAD is diagnosed when stenosis exceeds 50\% in an artery as evaluated by coronary angiography, which is the invasive gold standard test in this data set. The data set contains the following variables:

\begin{description}
\item[SPECT thallium test, $T$:] Binary, 1 = Positive, 0 = Negative
\item[CAD, $D$:] Binary, 1 = Yes, 0 = No
\item[Covariates, $\mathbf{X}$:]
\begin{enumerate}
\item[]
\item \textbf{Gender}, $X1$: Binary, 1 = Male, 0 = Female
\item \textbf{Stress mode}, $X2$: Binary, 1 = Dipyridamole, 0 = Exercise. Dipyridamole is a medication for stress test when the patient is unable to exercise.
\item \textbf{Age}, $X3$: Binary, 1 = 60 years and above, 0 = Below 60 years
\end{enumerate}
\end{description}
SPECT thallium test was performed on 2688 patients. Only 471 patients underwent coronary angiography for verification of the CAD status. The rest of the patients were unverified (2217 patients, 82.5\%).

\subsection{R environment and packages}
The code was tested using R version 3.6.3\cite{rcore} and RStudio version 1.4.1717.\cite{rstudio} A newly developed R package, namely \emph{PVBcorrect} version 0.1.0,\cite{pvbcorrect} was used. Two R packages, \emph{boot} version 1.3-28\cite{boot} and \emph{mice} version 3.13.0\cite{mice} were used indirectly via \emph{PVBcorrect}. R code and outputs in this article are differentiated by hashes, where R outputs are formatted with preceding hashes \#\#.

\subsection{Preliminaries}

\subsubsection{Install and load packages}
The \emph{PVBcorrect} R package was used in this tutorial. The package is available from GitHub at \url{https://github.com/wnarifin/PVBcorrect} and it can be easily installed using \emph{devtools} package as follows,

\begin{verbatim}
install.packages("devtools")
devtools::install_github("wnarifin/PVBcorrect")
\end{verbatim}
Then, the \emph{PVBcorrect} package can be loaded as follows,
\begin{verbatim}
library(PVBcorrect)
\end{verbatim}

\subsubsection{Data exploration}
The data set described in Subsection \ref{data_desc} is included in the \emph{PVBcorrect} package as a built-in data set. The data set was named \texttt{cad\_pvb} and its information can be viewed by

\begin{verbatim}
?cad_pvb
\end{verbatim}
The data set consists of 2688 observations and five categorical variables,
\begin{verbatim}
str(cad)
\end{verbatim}
\begin{verbatim}
## 'data.frame':    2688 obs. of  5 variables:
##  $ X1: int  0 0 0 0 0 0 0 0 0 0 ...
##  $ X2: int  0 0 0 0 0 0 0 0 0 0 ...
##  $ X3: int  0 0 0 0 0 0 0 0 0 0 ...
##  $ T : int  0 0 0 0 0 0 0 0 1 1 ...
##  $ D : int  0 0 0 0 0 0 0 1 0 0 ...
\end{verbatim}
The variable \texttt{D} (disease status, i.e. having CAD) contains 2217 missing observations,
\begin{verbatim}
summary(cad_pvb)
\end{verbatim}
\begin{verbatim}
##        X1               X2              X3               T         
##  Min.   :0.0000   Min.   :0.000   Min.   :0.0000   Min.   :0.0000  
##  1st Qu.:0.0000   1st Qu.:0.000   1st Qu.:0.0000   1st Qu.:0.0000  
##  Median :1.0000   Median :0.000   Median :0.0000   Median :1.0000  
##  Mean   :0.5737   Mean   :0.394   Mean   :0.4226   Mean   :0.5294  
##  3rd Qu.:1.0000   3rd Qu.:1.000   3rd Qu.:1.0000   3rd Qu.:1.0000  
##  Max.   :1.0000   Max.   :1.000   Max.   :1.0000   Max.   :1.0000  
##                                                                    
##        D         
##  Min.   :0.0000  
##  1st Qu.:0.0000  
##  Median :0.0000  
##  Mean   :0.4246  
##  3rd Qu.:1.0000  
##  Max.   :1.0000  
##  NA's   :2217
\end{verbatim}

\subsection{Selected methods}
In this section, an overview of each selected method is given. The overview is followed by applying the methods in a step-by-step tutorial using the \emph{PVBcorrect} R package. We purposely selected the BG method,\cite{begg1983} MI method\cite{harel2006} and EM-based logistic regression method\cite{kosinski2003a} because they are commonly used, cited and compared as the baseline methods described in the previous overview section.

\subsubsection{Complete case analysis (uncorrected method)}
As a baseline uncorrected method, the complete case analysis (CCA) method accuracy estimates are calculated from the complete cases only.\cite{degroot2011b} CCA is unbiased whenever the missing data mechanism is MCAR\citep{cecil1996, kosinski2003a, unal2014} (i.e. no partial verification bias), but it is biased in the presence of partial verification bias. CCA estimates of sensitivity (Se), specificity (Sp), PPV and NPV are obtained as follows:

\begin{equation}
\label{eq:ccasn}
\begin{split}
\widehat{Se}_{CCA} & = \hat{P}(T=1|D=1) \\ 
& = \frac{s_1}{s_1+s_0}
\end{split}
\end{equation}

\begin{equation}
\label{eq:ccasp}
\begin{split}
\widehat{Sp}_{CCA} & = \hat{P}(T=0|D=0) \\
& = \frac{r_0}{r_1+r_0}
\end{split}
\end{equation}

\begin{equation}
\label{eq:ccappv}
\begin{split}
\widehat{PPV}_{CCA} & = \hat{P}(D=1|T=1) \\ 
& = \frac{s_1}{s_1+r_1}
\end{split}
\end{equation}

\begin{equation}
\label{eq:ccanpv}
\begin{split}
\widehat{NPV}_{CCA} & = \hat{P}(D=0|T=0) \\ 
& = \frac{r_0}{s_0+r_0}
\end{split}
\end{equation}

\noindent The 100(1-$\alpha$)\% confidence interval (CI) of the CCA accuracy estimates can be calculated using Wald interval,

\begin{equation}
\label{eq:ccasn_ci}
\begin{split}
\widehat{Se}_{CCA} & \pm z_{1-\alpha/2} \sqrt{\frac{\widehat{Se}_{CCA}(1-\widehat{Se}_{CCA})}{s_1+s_0}}\\
& \pm z_{1-\alpha/2} \sqrt{\frac{s_1 s_0}{(s_1+s_0)^3}}
\end{split}
\end{equation}

\begin{equation}
\label{eq:ccasp_ci}
\begin{split}
\widehat{Sp}_{CCA} & \pm z_{1-\alpha/2} \sqrt{\frac{\widehat{Sp}_{CCA}(1-\widehat{Sp}_{CCA})}{r_1+r_0}}\\
& \pm z_{1-\alpha/2} \sqrt{\frac{r_1 r_0}{(r_1+r_0)^3}}
\end{split}
\end{equation}

\begin{equation}
\label{eq:ccappv_ci}
\begin{split}
\widehat{PPV}_{CCA} & \pm z_{1-\alpha/2} \sqrt{\frac{\widehat{PPV}_{CCA}(1-\widehat{PPV}_{CCA})}{s_1+r_1}}\\
& \pm z_{1-\alpha/2} \sqrt{\frac{s_1 r_1}{(s_1+r_1)^3}}
\end{split}
\end{equation}

\begin{equation}
\label{eq:ccanpv_ci}
\begin{split}
\widehat{NPV}_{CCA} & \pm z_{1-\alpha/2} \sqrt{\frac{\widehat{NPV}_{CCA}(1-\widehat{NPV}_{CCA})}{s_0+r_0}}\\
& \pm z_{1-\alpha/2} \sqrt{\frac{s_0 r_0}{(s_0+r_0)^3}}
\end{split}
\end{equation}

We first consider the cross-classification table of test result against disease status in the data set for the CCA method. This can be obtained easily by using the \texttt{view\_table} function in the \emph{PVBcorrect} package,

\begin{verbatim}
view_table(data = cad_pvb, test = "T", disease = "D")
\end{verbatim}
\begin{verbatim}
##      Disease
## Test  yes  no
##   yes 195 232
##   no    5  39
\end{verbatim}excluding the unverified cases. Using the \texttt{acc\_cca} function, we can obtain the CCA accuracy estimates and the respective 95\% CI,

\begin{verbatim}
cca_out = acc_cca(data = cad_pvb, test = "T", disease = "D", ci = TRUE)
cca_est = cca_out$acc_results
cca_est
\end{verbatim}
\begin{verbatim}
## Estimates of accuracy measures
## Uncorrected for PVB: Complete Case Analysis
##           Est         SE     LowCI     UppCI
## Sn  0.9750000 0.01103970 0.9533626 0.9966374
## Sp  0.1439114 0.02132173 0.1021216 0.1857013
## PPV 0.4566745 0.02410569 0.4094282 0.5039207
## NPV 0.8863636 0.04784519 0.7925888 0.9801385
\end{verbatim}

\subsubsection{Begg and Greenes' method}
Begg and Greenes\cite{begg1983} proposed a correction method based on Bayes' theorem (BG method) whenever the missing data mechanism is MAR. In a simple situation where there is no covariate, sensitivity and specificity estimates of the BG method are as follows\cite{begg1983, zhou1993, harel2006, zhou2011}:

\begin{equation}
\label{eq:bgsn}
\begin{split}
\widehat{Se}_{BG} & = \hat{P}(T=1|D=1) \\ 
& = \frac{\hat{P}(T=1)\hat{P}(D=1|T=1,V=1)}{\hat{P}(T=1)\hat{P}(D=1|T=1,V=1) + \hat{P}(T=0)\hat{P}(D=1|T=0,V=1)} \\ 
& = \frac{n_1s_1/(s_1+r_1)}{n_1s_1/(s_1+r_1) + n_0s_0/(s_0+r_0)}
\end{split}
\end{equation}

\begin{equation}
\label{eq:bgsp}
\begin{split}
\widehat{Sp}_{BG} & = \hat{P}(T=0|D=0) \\
& = \frac{\hat{P}(T=0)\hat{P}(D=0|T=0,V=1)}{\hat{P}(T=1)\hat{P}(D=0|T=1,V=1) + \hat{P}(T=0)\hat{P}(D=0|T=0,V=1)} \\
& = \frac{n_0r_0/(s_0+r_0)}{n_1r_1/(s_1+r_1) + n_0r_0/(s_0+r_0)}
\end{split}
\end{equation}

\noindent The 100(1-$\alpha$)\% CI of these BG estimates are given by

\begin{equation}
\label{eq:bgsn_ci}
\begin{split}
\widehat{Se}_{BG} & \pm z_{1-\alpha/2} \sqrt{[\widehat{Se}_{BG}(1-\widehat{Se}_{BG})]^2 \left(\frac{n}{n_0 n_1} + \frac{r_1}{s_1(s_1+r_1)} + \frac{r_0}{s_0(s_0+r_0)}\right)}
\end{split}
\end{equation}

\begin{equation}
\label{eq:bgsp_ci}
\begin{split}
\widehat{Sp}_{BG} & \pm z_{1-\alpha/2} \sqrt{[\widehat{Sp}_{BG}(1-\widehat{Sp}_{BG})]^2 \left(\frac{n}{n_0 n_1} + \frac{s_1}{r_1(s_1+r_1)} + \frac{s_0}{r_0(s_0+r_0)}\right)}
\end{split}
\end{equation}

Specific formulas to calculate PPV and NPV estimates of the BG method (no covariate) were given by de Groot et al., which they referred as B\&G1.\cite{degroot2011b} However, Zhou\cite{zhou1994effect, zhou1998} showed that when the MAR assumption holds, the BG estimates of PPV and NPV and the respective standard errors are the same as the CCA estimates.

\begin{equation}
\label{eq:bgppv}
\begin{split}
\widehat{PPV}_{BG} & = \widehat{PPV}_{CCA} = \hat{P}(D=1|T=1) \\ 
& = \frac{s_1}{s_1+r_1}
\end{split}
\end{equation}

\begin{equation}
\label{eq:bgnpv}
\begin{split}
\widehat{NPV}_{BG} & = \widehat{NPV}_{CCA} = \hat{P}(D=0|T=0) \\ 
& = \frac{r_0}{s_0+r_0}
\end{split}
\end{equation}Therefore, the calculation of 100(1-$\alpha$)\% CIs for PPV and NPV also follows the calculation for the CCA method (Equations \ref{eq:ccappv_ci} and \ref{eq:ccanpv_ci}).

In the presence of covariates, it is easier to use statistical models, such as the logistic regression model,\cite{begg1983} to estimate $P(D|T,\mathbf{X})$ and $P(T,\mathbf{X})$. Alonzo\cite{alonzo2005} provided an extension to Begg and Greenes' method\cite{begg1983} (EBG method), in which both categorical and continuous covariates can be included in the estimation. Although the EBG method was originally meant for a continuous index test, the method also applied to the binary index test.\cite{he2012} In addition, the EBG method reduces to the BG method when all covariates are categorical and applied using a saturated model,\cite{alonzo2005} in which all possible interaction terms between the index test and covariates are included. The sensitivity and specificity estimates of the EBG method are as follows \cite{alonzo2005, he2012}:

\begin{equation}
\label{eq:ebgsn}
\widehat{Se}_{EBG} = \frac{\sum_{i=1}^n T_i \hat{P}(D_i=1|T_i,\mathbf{X}_i)} {\sum_{i=1}^n \hat{P}(D_i=1|T_i,\mathbf{X}_i)}
\end{equation}

\begin{equation}
\label{eq:ebgsp}
\widehat{Sp}_{EBG} = \frac{\sum_{i=1}^n (1-T_i) [1-\hat{P}(D_i=1|T_i,\mathbf{X}_i)]} {\sum_{i=1}^n 1-\hat{P}(D_i=1|T_i,\mathbf{X}_i)}
\end{equation}
where $\hat{P}(D_i=1|T_i,\mathbf{X}_i)$ is obtained from a logistic regression of the observed data.

Using Bayes' theorem, PPV and NPV can be expressed as functions of sensitivity, specificity and disease prevalence as follows\cite{zhou2011}:

\begin{equation}
\label{eq:pvbayes}
\begin{split}
P(D=d|T=t) & = \frac{P(D=d)P(T=t|D=d)}{P(T=t)} \\
&= \frac{P(D=d)P(T=t|D=d)}{P(D=0)P(T=t|D=0) + P(D=1)P(T=t|D=1)}
\end{split}
\end{equation}

\noindent It follows that we can write PPV and NPV for the EBG method as

\begin{equation}
\label{eq:ebgppv}
\begin{split}
\widehat{PPV}_{EBG} & = \frac{\hat{P}(D=1)\hat{P}(T=1|D=1)}{\hat{P}(D=1)\hat{P}(T=1|D=1) + \hat{P}(D=0)\hat{P}(T=1|D=0)} \\
& = \frac{\hat{P}(D=1) \times \widehat{Se}_{EBG}}{\hat{P}(D=1) \times \widehat{Se}_{EBG} + \hat{P}(D=0) \times (1 - \widehat{Sp}_{EBG})} \\
& = \frac{\sum_{i=1}^n \hat{P}(D_i=1|T_i,\mathbf{X}_i) \times \widehat{Se}_{EBG}}{\sum_{i=1}^n  \hat{P}(D_i=1|T_i,\mathbf{X}_i) \times \widehat{Se}_{EBG} +  [1 - \hat{P}(D_i=1|T_i,\mathbf{X}_i)] \times (1 - \widehat{Sp}_{EBG})}
\end{split}
\end{equation}

\begin{equation}
\label{eq:ebgnpv}
\begin{split}
\widehat{NPV}_{EBG} & = \frac{\hat{P}(D=0)\hat{P}(T=0|D=0)}{\hat{P}(D=0)\hat{P}(T=0|D=0) + \hat{P}(D=1)\hat{P}(T=0|D=1)} \\
& = \frac{\hat{P}(D=0) \times \widehat{Sp}_{EBG}}{\hat{P}(D=0)  \times \widehat{Sp}_{EBG} + \hat{P}(D=1) \times (1 - \widehat{Se}_{EBG})} \\
& = \frac{\sum_{i=1}^n [1 - \hat{P}(D_i=1|T_i,\mathbf{X}_i)] \times \widehat{Sp}_{EBG}}
{\sum_{i=1}^n  [1 - \hat{P}(D_i=1|T_i,\mathbf{X}_i)] \times \widehat{Sp}_{EBG} + \hat{P}(D_i=1|T_i,\mathbf{X}_i) \times (1 - \widehat{Se}_{EBG})}
\end{split}
\end{equation}which give general PPV and NPV equations for the EBG method in the presence of covariates, where $\hat{P}(D_i=1|T_i,\mathbf{X}_i)$ is obtained from the logistic regression. These will reduce to Equations \ref{eq:bgppv} and \ref{eq:bgnpv} without covariate. Other equivalent formulas to calculate PPV and NPV estimates of the BG method for one covariate were given by de Groot et al. (referred to as B\&G2).\cite{degroot2011b}

For BG and EBG methods, the calculation of the standard errors and 95\% CI of the accuracy estimates gets quite complicated in the presence of covariates, and this can be referred to the original papers and other references.\cite{begg1983, zhou1993, alonzo2005, degroot2011b, zhou2011} However, it is easier to use bootstrapping to obtain the standard errors and 95\% CIs as implemented in other studies.\cite{alonzo2005, he2012, unal2014} Bootstrap samples are drawn from the verified cases, while the corresponding predicted probability values are obtained from all cases. Bootstrap bias-corrected and accelerated (BCa) interval was chosen for this tutorial as it is generally regarded as the method of choice\cite{woodward2014} and is available in the \emph{boot} R package.

Having reviewed the BG and EBG methods, now we can perform the analysis. We can obtain the cross-classification table of test result against disease status by

\begin{verbatim}
view_table(data = cad_pvb, test = "T", disease = "D", show_unverified = TRUE, show_total = TRUE)
\end{verbatim}
\begin{verbatim}
##      Disease
## Test   yes   no unverified Total
##   yes  195  232        996  1423
##   no     5   39       1221  1265
\end{verbatim}which now includes the unverified cases and total. This is similar to the typical data table shown before (Table \ref{tab:pvb}).

Using the \texttt{acc\_ebg} function, we can obtain the EBG accuracy estimates and the respective 95\% CI. Here, we need to set the number of bootstrap samples (replications), \texttt{R}, because bootstrapping is used to obtain the 95\% CI, where we set this option as \texttt{R = 999}. Using odd numbers, such as 999 can save computer time.\cite{woodward2014} The random seed number, \texttt{seednum} is set to 12345 to allow the readers to replicate the results throughout this article. For the reason mentioned before, we set the CI type, \texttt{ci\_type} as \texttt{"bca"}.

\begin{verbatim}
ebg_out = acc_ebg(data = cad_pvb, test = "T", disease = "D", ci = TRUE, ci_type = "bca",
                  seednum = 12345, R = 999)
ebg_est = ebg_out$acc_results
ebg_est
\end{verbatim}
\begin{verbatim}
## Estimates of accuracy measures
## Corrected for PVB: Extended Begg and Greenes' Method
##           Est         SE     LowCI     UppCI
## Sn  0.8188629 0.06438696 0.6808328 0.9305696
## Sp  0.5918754 0.01759285 0.5538231 0.6241082
## PPV 0.4566745 0.02421467 0.4090813 0.5083287
## NPV 0.8863636 0.04925696 0.7662900 0.9605648
\end{verbatim}The readers may also try the \texttt{acc\_bg} function that uses the CI formulas in Equations \ref{eq:bgsn_ci}, \ref{eq:bgsp_ci}, \ref{eq:ccappv_ci} and \ref{eq:ccanpv_ci}. The readers may notice minimal differences in the CI values between the methods.

Now we use the EBG method with a covariate $X3$,
\begin{verbatim}
ebgx_out = acc_ebg(data = cad_pvb, test = "T", disease = "D", covariate = "X3",
                   saturated_model = TRUE, ci = TRUE, ci_type = "bca",
                   seednum = 12345, R = 999)
ebgx_est = ebgx_out$acc_results
ebgx_est
\end{verbatim}
\begin{verbatim}
##           Est         SE     LowCI     UppCI
## Sn  0.8400495 0.06060738 0.6968409 0.9360124
## Sp  0.5912022 0.01573772 0.5566062 0.6194934
## PPV 0.4437285 0.02342825 0.3993989 0.4935086
## NPV 0.9049587 0.04311795 0.7883877 0.9669421
\end{verbatim}Setting the \texttt{saturated\_model} option as \texttt{TRUE} allows specifying the EBG to give an equivalent result to the BG method by utilising a saturated model. Also, note the differences in the estimates after including $X3$, most notable for sensitivity, PPV and NPV.

\subsubsection{MI}
Harel and Zhou \cite{harel2006} proposed using MI as an alternative to the BG method. Its main advantage is that it allows the use of complete data methods\cite{little2020} to obtain accuracy estimates and the respective confidence intervals.\cite{harel2006} MI is a simulation method to handle missing data\cite{schafer1997} involving repeated imputation of missing values by drawing the values from posterior predictive distribution.\cite{rubin1996, little2020} In the context of diagnostic accuracy studies, each missing disease status is replaced by $m>1$ plausible values, resulting in $m$ complete data sets.\cite{harel2006, alonzo2014} Each of these data sets is then analysed by complete data methods, thereafter the $m$ estimates are combined to provide final estimates.\citep{harel2006, alonzo2014}

Here, the imputation step of the MI method is demonstrated by logistic regression.\cite{harel2006} The number of imputed data sets ($m$) follows the rule of thumb, where $m$ equals the percentage of incomplete cases.\cite{bodner2008, white2011, Pedersen2017} Accuracy estimates for the MI method are straightforward, where the estimates are calculated according to the CCA method (Equations \ref{eq:ccasn}, \ref{eq:ccasp}, \ref{eq:ccappv} and \ref{eq:ccanpv}) for each imputed data set. The mean of the accuracy estimates, calculated from these $m$ imputed data sets, are then taken as MI estimates of these accuracy measures:

\begin{equation}
\label{eq:misn}
\overline{Se}_{MI} = \frac{1}{m} \sum_{j=1}^m \widehat{Se}_{CCA, j}
\end{equation}

\begin{equation}
\label{eq:misp}
\overline{Sp}_{MI} = \frac{1}{m} \sum_{j=1}^m \widehat{Sp}_{CCA, j}
\end{equation}

\begin{equation}
\label{eq:mippv}
\overline{PPV}_{MI} = \frac{1}{m} \sum_{j=1}^m \widehat{PPV}_{CCA, j}
\end{equation}

\begin{equation}
\label{eq:minpv}
\overline{NPV}_{MI} = \frac{1}{m} \sum_{j=1}^m \widehat{NPV}_{CCA, j}
\end{equation}

For the estimates obtained by MI, the calculation of standard errors and 95\% CIs for sensitivity and specificity can be performed by following Rubin's rule.\citep{rubin1996,van2018,little2020} Let $\hat{Q}^{(m)}$ be the point estimate (i.e. sensitivity, specificity, PPV or NPV) and $\hat{U}^{(m)}$ be the variance estimate for each of $m$th set of imputed data, $m=1,2,...,M$. The MI point estimate $\bar{Q}$ is calculated for each accuracy measure as shown in Equations \ref{eq:misn}, \ref{eq:misp}, \ref{eq:mippv} and \ref{eq:minpv}. The variability associated with $\bar{Q}$ has two components, which are the within-imputation variance

\begin{equation}
\label{eq:ubarm}
\bar{U} = \frac{1}{M}\sum_{m=1}^M \hat{U}^{(m)}
\end{equation}
and the between-imputation variance
\begin{equation}
\label{eq:bm}
B = \frac{1}{M-1}\sum_{m=1}^M (\hat{Q}^{(m)}-\bar{Q})^2
\end{equation}
giving a total variability associated with $\bar{Q}$
\begin{equation}
\label{eq:tm}
T = \bar{U} + \left(1+\frac{1}{M}\right)B
\end{equation}
where $(1+1/M)$ is an adjustment for small $M$.

The variance estimates of sensitivity, specificity, PPV and NPV for each $m$th set of imputed data $\hat{U}^{(m)}$ are calculated as

\begin{equation}
\label{eq:uhatsn}
\hat{U}^{(m)}_{Se} = \frac{s_1 s_0}{(s_1+s_0)^3}
\end{equation}

\begin{equation}
\label{uhatsp}
\hat{U}^{(m)}_{Sp} = \frac{r_0 r_1}{(r_1+r_0)^3}
\end{equation}

\begin{equation}
\label{eq:uhatppv}
\hat{U}^{(m)}_{PPV} = \frac{s_1 r_1}{(s_1+r_1)^3}
\end{equation}

\begin{equation}
\label{eq:uhatnpv}
\hat{U}^{(m)}_{NPV} = \frac{s_0 r_0}{(s_0+r_0)^3}
\end{equation}

Note that the quantity $s_1 + s_0$ and $r_1 + r_0$ will be different for each $m$th set of imputed data, because these are marginal counts by $D=1$ and $D=0$. The within-imputation variances $\bar{U}_M$ of the accuracy measures for each $m$th set of imputed data are calculated as

\begin{equation}
\label{eq:ubarall}
\bar{U}_{Acc} = \frac{1}{M}\sum_{m=1}^M \hat{U}^{(m)}_{Acc}
\end{equation}where $Acc$ refers to each of the accuracy measures, i.e. Se, Sp, PPV and NPV.

The calculation of between-imputation variance $B$ and total variance $T$ of sensitivity and specificity is straightforward, as given in Equations \ref{eq:bm} and \ref{eq:tm}. The 100(1-$\alpha$)\% CI of the MI estimates of sensitivity and specificity are then calculated as follows:\cite{harel2006}

\begin{equation}
\label{eq:miall_ci}
\overline{Acc}_{MI} \pm t_{(v,1-\alpha/2)} \sqrt{T_{Acc}}
\end{equation}where the degrees of freedom, $v$ is calculated for each of the accuracy measures as follows:

\begin{equation}
\label{eq:dfmi}
v = (M-1)\left[1+\frac{\bar{U}}{(1+1/M)B}\right]^2
\end{equation}

The correction by MI method can be performed by using the \texttt{acc\_mi} function. By default, the function will use logistic regression for the imputation. The number of imputation \texttt{m} is set to 85 based on the percentage of missing observations\cite{white2011}

\begin{verbatim}
mean(is.na(cad_pvb$D)) * 100
\end{verbatim}
\begin{verbatim}
## [1] 82.47768
\end{verbatim}Again, the random seed, \texttt{seednum} is set to $12345$ to allow replication of the results. Now, we obtain the accuracy estimates with the 95\% CI by the MI method

\begin{verbatim}
mi_out = acc_mi(data = cad_pvb, test = "T", disease = "D", ci = TRUE, seednum = 12345, m = 85)
mi_est = mi_out$acc_results
mi_est
\end{verbatim}
\begin{verbatim}
## Estimates of accuracy measures
## Corrected for PVB: Multiple Imputation Method
##           Est         SE     LowCI     UppCI
## Sn  0.8026453 0.06747072 0.6686296 0.9366609
## Sp  0.5867365 0.02132904 0.5446224 0.6288506
## PPV 0.4573767 0.02472496 0.4085569 0.5061965
## NPV 0.8685236 0.06027250 0.7487484 0.9882988
\end{verbatim}

Next, we perform the correction by the MI method with a covariate $X3$. Here, $X3$ is included in the imputation model to generate the imputed values. Additionally, note the differences in the estimates with the inclusion of $X3$, most notable for sensitivity, PPV and NPV estimates.

\begin{verbatim}
mix_out = acc_mi(data = cad_pvb, test = "T", disease = "D", covariate = "X3", ci = TRUE, 
                 seednum = 12345, m = 85)
mix_est = mix_out$acc_results
mix_est
\end{verbatim}
\begin{verbatim}
## Estimates of accuracy measures
## Corrected for PVB: Multiple Imputation Method
##           Est         SE     LowCI     UppCI
## Sn  0.8175575 0.06014723 0.6981222 0.9369927
## Sp  0.5864481 0.01914576 0.5486909 0.6242052
## PPV 0.4455872 0.02344523 0.3993230 0.4918514
## NPV 0.8852732 0.04537927 0.7951304 0.9754160
\end{verbatim}

\subsubsection{EM-based logistic regression}\label{sec:emlogreg}
Konsinski and Barnhart\cite{kosinski2003a} proposed a method using logistic regression models with EM algorithm to correct for partial verification bias with MNAR missing data mechanism. The method is implemented by running multiple cycles of weighted logistic regression models on pseudo-data. The authors considered the selection model for joint probability of $V$, $T$ and $D$ given covariates $\mathbf{X}$ as $P(V,T,D|\mathbf{X}) = P(D|\mathbf{X}) \times P(T|D,\mathbf{X}) \times P(V|T,D,\mathbf{X})$. They parameterise the components in the equation with three separate logistic regression models:

\begin{enumerate}
\item \textbf{Disease component:} $logit\ P(D_i=1|\mathbf{x}_i)=\mathbf{z}_{0i}^{'}\mathbf{\alpha}$
\item \textbf{Diagnostic test component:} $logit\ P(T_i=1|D_i,\mathbf{x}_i)=\mathbf{z}_{1i}^{'}\mathbf{\beta}$
\item \textbf{Missing data mechanism component:} $logit\ P(R_i=1|D_i,T_i,\mathbf{x}_i)=\mathbf{z}_{2i}^{'}\mathbf{\gamma}$
\end{enumerate}

\noindent where $logit(p) = log\{p/(1-p)\}$, $\mathbf{\theta} = (\mathbf{\alpha^{'}}, \mathbf{\beta^{'}}, \mathbf{\gamma^{'}})$ is the vector of parameters, and $\mathbf{z}_{mi}^{'}$ is the $i$th row ($i=1,...,n$) of the design matrix for the $m$th logistic regression model ($m=0,1,2$) with a choice of $D_i,T_i,\mathbf{x}_i$.

The correction method is implemented as follows:
\begin{enumerate}
\item Pseudo-data is created with $n+u$ rows. The first $n-u$ rows are for verified patients ($V=1$), while the remaining $2u$ rows are double-stacked data for unverified patients with missing disease status, $D$. The $D$ value for the first $u$ stack $D$ is set as $D=0$, and the second $u$ stack $D$ is set as $D=1$.
\item Weighted logistic regression models are fitted iteratively (M-step). Weights $w_k(\theta)$ are updated at each iteration in the preceding E-step.
\end{enumerate}

The weights $w_k(\theta)$ are defined as follows:
$$
w_{k}(\theta) = \begin{cases}
1 & \text{for}\ k=1,...,n-u\\
p_{k}(\theta)/\{p_k(\theta) + p_{k+U}(\theta) \} & \text{for}\ k=n-u+1,...,n\\
1-w_{k-u}(\theta) & \text{for}\ k=n+1,...,n+u
\end{cases}
$$
\noindent where $p_{k}(\theta) = \prod_{m=0}^2 \{p_{mk}(\theta)\}^{y_{mk}} \{1-p_{mk}(\theta)\}^{1-y_{mk}}$. $p_{mk}$ can be easily obtained from the respective logistic regression models for each observation at each iteration using standard logistic regression module by obtaining the probability of the outcomes (i.e. $D=1$, $T=1$ or $V=1$) for each of the patients.

By using the coefficients from the fitted logistic regression model for the diagnostic test component (without covariate), sensitivity and specificity estimates are easily calculated by obtaining the predicted probabilities for $D=1$ and $D=0$,

\begin{equation}
\label{eq:kossn}
\begin{split}
\widehat{Se}_{EM} & = \hat{P}(T=1|D=1)
\end{split}
\end{equation}

\begin{equation}
\label{eq:kosgsp}
\begin{split}
\widehat{Sp}_{EM} & = \hat{P}(T=0|D=0) \\
& = 1 - \hat{P}(T=1|D=0)
\end{split}
\end{equation}

Although the calculation of PPV and NPV estimates were not provided in the original paper by Kosinski and Barnhart,\cite{kosinski2003a} the estimates can be calculated using Bayes' theorem, where PPV and NPV are expressed as functions of sensitivity, specificity and disease prevalence.\cite{zhou2011}

\begin{equation}
\label{eq:kosppv}
\begin{split}
\widehat{PPV}_{EM} & = \frac{\hat{P}(D=1)\hat{P}(T=1|D=1)}{\hat{P}(D=1)\hat{P}(T=1|D=1) + \hat{P}(D=0)\hat{P}(T=1|D=0)} \\
& = \frac{\hat{P}(D=1) \times \widehat{Se}_{EM}}{\hat{P}(D=1) \times \widehat{Se}_{EM} + [1 - \hat{P}(D=1)] \times (1 - \widehat{Sp}_{EM})}
\end{split}
\end{equation}

\begin{equation}
\label{eq:kosnpv}
\begin{split}
\widehat{NPV}_{EM} & = \frac{\hat{P}(D=0)\hat{P}(T=0|D=0)}{\hat{P}(D=0)\hat{P}(T=0|D=0) + \hat{P}(D=1)\hat{P}(T=0|D=1)} \\
& = \frac{[1 - \hat{P}(D=1)] \times \widehat{Sp}_{EM}}{[1 - \hat{P}(D=1)]  \times \widehat{Sp}_{EM} + \hat{P}(D=1) \times (1 - \widehat{Se}_{EM})}
\end{split}
\end{equation}where $\hat{P}(D=1)$ is the predicted probability obtained from the fitted logistic regression model for the disease component, which is an intercept only model in the absence of covariate.

For the model with covariates $\mathbf{X}$, the marginal sensitivity and specificity estimates can be calculated for covariate patterns of categorical $\mathbf{X}$ as follows\cite{kosinski2003a, buzoianu2008}:
\begin{equation}
\label{eq:kossn1}
\begin{split}
\widehat{Se} & = \hat{P}(T=1|D=1) \\
& = \frac {\hat{P}(T=1,D=1)} {\hat{P}(D=1)} \\
& = \frac {\sum_\mathbf{X}\hat{P}(T=1|D=1,\mathbf{X})\hat{P}(D=1|\mathbf{X})w_\mathbf{X}} {\sum_\mathbf{X}\hat{P}(D=1|\mathbf{X})w_\mathbf{X}}
\end{split}
\end{equation}

\begin{equation}
\label{eq:kossp1}
\begin{split}
\widehat{Sp} & = \hat{P}(T=0|D=0) \\
& = \frac {\hat{P}(T=0,D=0)} {\hat{P}(D=0)} \\
& = \frac {\sum_\mathbf{X}\hat{P}(T=0|D=0,\mathbf{X})\hat{P}(D=0|\mathbf{X})w_\mathbf{X}} {\sum_\mathbf{X}\hat{P}(D=0|\mathbf{X})w_\mathbf{X}} \\
& = \frac {\sum_\mathbf{X}[1 - \hat{P}(T=1|D=0,\mathbf{X})][1 - P(D=1|\mathbf{X})]w_\mathbf{X}} {\sum_\mathbf{X}[1 - \hat{P}(D=1|\mathbf{X})]w_\mathbf{X}}
\end{split}
\end{equation}
where $w_\mathbf{X}$ is the population weight of covariate pattern $\mathbf{X}$. This can be implemented by using the weights of covariate patterns from the sample $n$.

By employing the method described in Muller and MacLehose\cite{muller2014} through setting $D=1$ to all observations $n$, the sensitivity estimate can be calculated by obtaining predicted probabilities on this modified data set (all observations set as $D=1$) using the coefficients from the fitted logistic regression models. Equation \ref{eq:kossn1} can also be modified to include continuous covariates. Therefore, the sensitivity estimate can be calculated as follows:
\begin{equation}
\label{eq:kossn1a}
\begin{split}
\widehat{Se}_{EM} & = \hat{P}(T=1|D=1) \\
& = \frac {\hat{P}(T=1,D=1)} {\hat{P}(D=1)} \\
& = \frac{\sum_{i=1}^n \hat{P}(T_i=1|D_i=1,\mathbf{x}_i)\hat{P}(D_i=1|\mathbf{x}_i)} {\sum_{i=1}^n \hat{P}(D_i=1|\mathbf{x}_i)} \\
& = \frac{\sum_{i=1}^n p_{2i} p_{1i}}{\sum_{i=1}^n p_{1i}}
\end{split}
\end{equation}

The same method\cite{muller2014} is performed by setting $D=0$ to all observations $n$. The specificity estimate can be calculated by obtaining predicted probabilities on this modified data set (all observations set as $D=0$) using the coefficients from the fitted logistic regression models. Equation \ref{eq:kossp1} can be modified to include continuous covariates; thus, the specificity estimate can be calculated as follows:
\begin{equation}
\label{eq:kossp1a}
\begin{split}
\widehat{Sp}_{EM} & = \hat{P}(T=0|D=0) \\
& = \frac {\hat{P}(T=0,D=0)} {\hat{P}(D=0)} \\
& = \frac{\sum_{i=1}^n \hat{P}(T_i=0|D_i=0,\mathbf{x}_i)\hat{P}(D_i=0|\mathbf{x}_i)} {\sum_{i=1}^n \hat{P}(D_i=0|\mathbf{x}_i)} \\
& = \frac{\sum_{i=1}^n [1-\hat{P}(T_i=1|D_i=0,\mathbf{x}_i)][1-\hat{P}(D_i=1|\mathbf{x}_i)]} {\sum_{i=1}^n [1-\hat{P}(D_i=1|\mathbf{x}_i)]} \\
& = \frac{\sum_{i=1}^n (1 - p_{2i}) (1 - p_{1i})} {\sum_{i=1}^n (1 - p_{1i})}
\end{split}
\end{equation}

Then, it follows that in the presence of covariates, the PPV and NPV estimates for the EM method can be written as
\begin{equation}
\label{eq:kosppv1a}
\begin{split}
\widehat{PPV}_{EM} & = \frac{\hat{P}(D=1)\hat{P}(T=1|D=1)}{\hat{P}(D=1)\hat{P}(T=1|D=1) + \hat{P}(D=0)\hat{P}(T=1|D=0)} \\
& = \frac{\hat{P}(D=1) \times \widehat{Se}_{EM}}{\hat{P}(D=1) \times \widehat{Se}_{EM} + [1 - \hat{P}(D=1)] \times (1 - \widehat{Sp}_{EM})} \\
& = \frac{\sum_{i=1}^n \hat{P}(D_i=1|\mathbf{x}_i) \times \widehat{Se}_{EM}}{\sum_{i=1}^n \hat{P}(D_i=1|\mathbf{x}_i) \times \widehat{Se}_{EM} + [1 - \hat{P}(D_i=1|\mathbf{x}_i)] \times (1 - \widehat{Sp}_{EM})} \\
& = \frac{\sum_{i=1}^n p_{1i} \times \widehat{Se}_{EM}}{\sum_{i=1}^n p_{1i} \times \widehat{Se}_{EM} + (1 - p_{1i}) \times (1 - \widehat{Sp}_{EM})}
\end{split}
\end{equation}

\begin{equation}
\label{eq:kosnpv1a}
\begin{split}
\widehat{NPV}_{EM} & = \frac{\hat{P}(D=0)\hat{P}(T=0|D=0)}{\hat{P}(D=0)\hat{P}(T=0|D=0) + \hat{P}(D=1)\hat{P}(T=0|D=1)} \\
& = \frac{[1 - \hat{P}(D=1)] \times \widehat{Sp}_{EM}}{[1 - \hat{P}(D=1)]  \times \widehat{Sp}_{EM} + \hat{P}(D=1) \times (1 - \widehat{Se}_{EM})} \\
& = \frac{\sum_{i=1}^n [1-\hat{P}(D_i=1|\mathbf{x}_i)] \times \widehat{Sp}_{EM}}
{\sum_{i=1}^n [1-\hat{P}(D_i=1|\mathbf{x}_i)]  \times \widehat{Sp}_{EM} + \hat{P}(D_i=1|\mathbf{x}_i) \times (1 - \widehat{Se}_{EM})} \\
& = \frac{\sum_{i=1}^n (1-p_{1i}) \times \widehat{Sp}_{EM}}
{\sum_{i=1}^n (1-p_{1i})  \times \widehat{Sp}_{EM} + p_{1i} \times (1 - \widehat{Se}_{EM})}
\end{split}
\end{equation}

Readers are encouraged to repeat the analysis with all three covariates available in the data set (set a higher \texttt{max\_t} value for the analysis). The results can be verified with the results presented in the original paper.\cite{kosinski2003a}

For the EM-based method, calculating the standard errors and 95\% CI for the accuracy estimates is complicated, which can be referred to in the original paper.\cite{kosinski2003a} This is easier to obtain by bootstrapping, as demonstrated by {\"U}nal and Burgut.\cite{unal2014} Bootstrap samples are drawn from the verified cases, followed by creating pseudo-data as described in Section \ref{sec:emlogreg} for each bootstrap sample. The EM algorithm is then run on each of these samples to obtain the bootstrapped CIs of the accuracy estimates. Like the EBG method, bootstrap bias-corrected and accelerated (BCa) interval is used in this tutorial.

Using the \texttt{acc\_em} function, we can obtain the EM accuracy estimates and the respective 95\% CI. In the same way that we use the EBG function \texttt{acc\_ebg}, we must set the seed number, number of bootstrap replications and CI type. In addition to these options, there are two specific options for EM; the maximum number of EM iteration, \texttt{t\_max} and the cutoff value for the minimum change between EM iteration, \texttt{cutoff}. This cutoff value defines the convergence of the EM procedure. For this tutorial, we set these values as \texttt{t\_max = 5000} and \texttt{cutoff = 0.0002}. We start with the EM method without covariate,

\begin{verbatim}
em_out = acc_em(data = cad_pvb, test = "T", disease = "D", ci = TRUE, ci_type = "bca",
                seednum = 12345, R = 999,
                t_max = 5000, cutoff = 0.0002)
em_est = em_out$acc_results
em_est
\end{verbatim}
\begin{verbatim}
## Estimates of accuracy measures
## Corrected for PVB: EM-based Method
##           Est         SE     LowCI     UppCI
## Sn  0.7123288 0.04749658 0.6215704 0.8014492
## Sp  0.6441335 0.04537897 0.5568974 0.7307390
## PPV 0.6550121 0.04424218 0.5735726 0.7398906
## NPV 0.7024365 0.04908850 0.6097175 0.7980444
\end{verbatim}Please note that the execution of the code will take some time to finish (in hours). So, the readers might want to start by running the code with a smaller bootstrap samples number, for example \texttt{R = 99}.

Next, we perform the EM method with a covariate $X3$. The \texttt{t\_max} is increased to 50,000 as it will take more iterations until convergence. Similarly, please note that the execution of this code will also take some time to finish. In our experience, it took us 15h with a GPU-accelerated numerical computation library (NVBLAS), so it can take longer for a typical CPU-based numerical computation.
\begin{verbatim}
emx_out = acc_em(data = cad_pvb, test = "T", disease = "D", covariate = "X3",
                 ci = TRUE, ci_type = "bca",
                 seednum = 12345, R = 999,
                 t_max = 50000, cutoff = 0.0002)
emx_est = emx_out$acc_results
emx_est
\end{verbatim}
\begin{verbatim}
## Estimates of accuracy measures
## Corrected for PVB: EM-based Method
##           Est         SE      LowCI     UppCI
## Sn  0.7131379 0.09978819 0.51169404 0.8919660
## Sp  0.6457868 0.06415708 0.54668184 0.8162199
## PPV 0.6574623 0.15285095 0.32472999 0.8369435
## NPV 0.7025015 0.17875084 0.08851609 0.9349706
\end{verbatim}

\subsection{Comparison of results}
The analysis results using the described methods can be easily summarised in a table format, as shown in Table \ref{tab:res1}. The R code for the table is given below (\emph{tibble} and \emph{knitr} R packages are required).
\begin{verbatim}
tbl_combined = tibble::tibble(  # Combine all results into a tibble
  Estimates = c("Sensitivity", "SE", "2.5%", "97.5%", "Specificity", "SE", "2.5%", "97.5%", 
                "PPV", "SE", "2.5%", "97.5%", "NPV", "SE", "2.5%", "97.5%"),
  CCA = c(t(cca_est["Sn", ]), t(cca_est["Sp", ]), t(cca_est["PPV", ]), t(cca_est["NPV", ])),
  EBG = c(t(ebg_est["Sn", ]), t(ebg_est["Sp", ]), t(ebg_est["PPV", ]), t(ebg_est["NPV", ])),
  EBGX = c(t(ebgx_est["Sn", ]), t(ebgx_est["Sp", ]), t(ebgx_est["PPV", ]), t(ebgx_est["NPV", ])),
  MI = c(t(mi_est["Sn", ]), t(mi_est["Sp", ]), t(mi_est["PPV", ]), t(mi_est["NPV", ])),
  MIX = c(t(mix_est["Sn", ]), t(mix_est["Sp", ]), t(mix_est["PPV", ]), t(mix_est["NPV", ])),
  EM = c(t(em_est["Sn", ]), t(em_est["Sp", ]), t(em_est["PPV", ]), t(em_est["NPV", ])),
  EMX = c(t(emx_est["Sn", ]), t(emx_est["Sp", ]), t(emx_est["PPV", ]), t(emx_est["NPV", ]))
); tbl_combined
# Round the results to 3 decimal places for viewing
tbl_combined_rnd = tbl_combined
tbl_combined_rnd[-1] = round(tbl_combined[-1], 3)
# View using knitr::kable
tbl_view = knitr::kable(tbl_combined_rnd, format = "simple", 
                        caption = "Comparison of accuracy estimates by PVB correction methods.")
  # format = "simple", html", "latex", "pipe", "rst"
tbl_view
\end{verbatim}

\begin{center}
\begin{table}[t]%
\centering
\caption{Comparison of accuracy estimates and their respective SEs and 95\% CIs by PVB correction methods.\label{tab:res1}}%

\begin{tabular*}{500pt}{@{\extracolsep\fill}lrrrrrrr@{\extracolsep\fill}}
\toprule
\textbf{Estimates} & \textbf{CCA} & \textbf{EBG} & \textbf{EBGX} & \textbf{MI} & \textbf{MIX} & \textbf{EM} & \textbf{EMX}\\
\midrule
\textbf{Sensitivity} & 0.975 & 0.819 & 0.840 & 0.803 & 0.818 & 0.712 & 0.713\\
SE & 0.011 & 0.064 & 0.061 & 0.067 & 0.060 & 0.047 & 0.100\\
2.5\% & 0.953 & 0.681 & 0.697 & 0.669 & 0.698 & 0.622 & 0.512\\
97.5\% & 0.997 & 0.931 & 0.936 & 0.937 & 0.937 & 0.801 & 0.892\\
\midrule
\textbf{Specificity} & 0.144 & 0.592 & 0.591 & 0.587 & 0.586 & 0.644 & 0.646\\
SE & 0.021 & 0.018 & 0.016 & 0.021 & 0.019 & 0.045 & 0.064\\
2.5\% & 0.102 & 0.554 & 0.557 & 0.545 & 0.549 & 0.557 & 0.547\\
97.5\% & 0.186 & 0.624 & 0.619 & 0.629 & 0.624 & 0.731 & 0.816\\
\midrule
\textbf{PPV} & 0.457 & 0.457 & 0.444 & 0.457 & 0.446 & 0.655 & 0.657\\
SE & 0.024 & 0.024 & 0.023 & 0.025 & 0.023 & 0.044 & 0.153\\
2.5\% & 0.409 & 0.409 & 0.399 & 0.409 & 0.399 & 0.574 & 0.325\\
97.5\% & 0.504 & 0.508 & 0.494 & 0.506 & 0.492 & 0.740 & 0.837\\
\midrule
\textbf{NPV} & 0.886 & 0.886 & 0.905 & 0.869 & 0.885 & 0.702 & 0.703\\
SE & 0.048 & 0.049 & 0.043 & 0.060 & 0.045 & 0.049 & 0.179\\
2.5\% & 0.793 & 0.766 & 0.788 & 0.749 & 0.795 & 0.610 & 0.089\\
97.5\% & 0.980 & 0.961 & 0.967 & 0.988 & 0.975 & 0.798 & 0.935\\
\bottomrule
\end{tabular*}

\begin{tablenotes}
\item Abbreviations: CCA, complete case analysis; CI, confidence interval; EBG, extended Begg and Greenes' method; EBGX, extended Begg and Greenes' method with a covariate; EM, EM-based logistic regression method (MNAR); EMX, EM-based logistic regression method (MNAR) with a covariate; MI, multiple imputation (logistic regression); MIX, multiple imputation (logistic regression) with a covariate; MNAR, missing not at random; NPV, negative predictive value; PPV, positive predictive value; PVB, partial verification bias; SE, standard error.
\end{tablenotes}

\end{table}
\end{center}

According to the results, the sensitivity and specificity estimates using the CCA method were notably biased. The results of the EBG method and MI method are presented for the MAR missing data mechanism, where the estimates for these methods differed slightly. The results of the EM method are presented for the MNAR missing data mechanism. Readers can change the missing data mechanism to MAR in the code by including \texttt{MNAR = FALSE} option to \texttt{acc\_em()} function so that verification \emph{V} does not depend on \emph{D}. The method will give similar results to the BG-based methods under the MAR assumption.

\section{Concluding remarks}\label{sec4}

This article provides a brief overview of the existing methods to correct for partial verification bias involving binary tests in diagnostic accuracy studies. The overview was then followed with a tutorial on implementing three selected methods in R using the \emph{PVBcorrect} R package. It is hoped that the overview, supplemented by the practical implementation in R, will make the methods more understandable and accessible to diagnostic test developers and researchers. The \emph{PVBcorrect} R package is available for installation from GitHub, which will make the methods more accessible to the researchers.


\section*{Acknowledgements}
We thank our colleagues at the School of Computer Sciences and the School of Medical Sciences, Universiti Sains Malaysia for the comments on this article's draft and the \emph{PVBcorrect} R package. We also thank two anonymous reviewers for their constructive comments, which greatly improved the article's readability.

\subsection*{Author contributions}
WNA and UKY conceived the experiments,  WNA conducted the experiments and analysed the results.  WNA and UKY prepared and reviewed the manuscript. 

\subsection*{Financial disclosure}
None reported.

\subsection*{Conflict of interest}
The authors declare no potential conflict of interests.

\section*{Supporting information}
The \emph{PVBcorrect} R package can be installed by following the instruction in this article or in the GitHub repository for the package at \url{https://github.com/wnarifin/PVBcorrect}. The R script containing all the code described in this article can be downloaded from \url{https://github.com/wnarifin/pvb-rtutorial}.

\bibliography{wileyNJD-AMA.bib}%

\end{document}